\documentclass[%
letterpaper,
reprint,
amsfonts,amsmath,amssymb,
aip,
jmp,
floatfix
]{revtex4-1}
\usepackage{graphicx}
\usepackage{dcolumn}
\usepackage{bm}

\usepackage{ulem}
\usepackage{xcolor}

\begin{document}
\title{Shell Model of {BaTiO$_3$} Derived from {\it{ab-initio}} Total Energy Calculations}

\author{J.~M.~Vielma}
\affiliation{Department of Physics, Oregon State University, Corvallis, Oregon 97331, USA}

\author{G.~Schneider}
\email{Guenter.Schneider@physics.oregonstate.edu}
\affiliation{Department of Physics, Oregon State University, Corvallis, Oregon 97331, USA}


\begin{abstract}
A shell model for ferroelectric perovskites fitted to results of first-principles 
density functional theory (DFT) calculations is strongly affected by approximations made in the 
exchange-correlation functional within DFT, and in general not as accurate as a 
shell model derived from experimental data.
We have developed an isotropic shell model for {BaTiO$_3$} based on the PBEsol 
exchange-correlation functional \cite{perdew_restoring_2008}, which
was specifically designed for crystal properties of solids.
Our shell model for {BaTiO$_3$} agrees with groundstate DFT properties and the experimental
lattice constants at finite temperatures.
The sequence of phases of {BaTiO$_3$} (rhombohedral, orthorhombic, tetragonal, 
cubic) is correctly reproduced but the temperature scale of the phase
transitions is compressed. The temperature scale can be improved by
scaling of the {\it{ab-initio}} energy surface. 
\end{abstract}

\maketitle

\section{Introduction}
\label{intro}
Ferroelectric perovskites or perovskite solid solution near a
ferroelectric instability have potential applications as fast
capacitive storage medium \cite{huang_phase_2008,ogihara_weakly_2009} and considerable 
interest exists to develop lead-free ferroelectric materials 
to replace the widely used piezoelectric Pb(Zr,Ti)O$_3$ (PZT).
\cite{zhang_lead_2008,wang_enhanced_2010}
Systematic selection and improvement of materials and properties for
applications depends on the detailed knowledge of the physics in these
materials, which frequently requires understanding of phenomena over
medium and long length and time scales.
Atomistic model simulations of ferroelectric perovskites, such
as isotropic \cite{sepliarsky_atomic-level_2000} and anisotropic
\cite{tinte_atomistic_1999} shell models
and approaches based on effective Hamiltonians \cite{rabe_first_1992},
can probe length and time scales that are beyond the reach of
computationally much more demanding first-principles simulations. 
In order to observe finite temperature phase transitions, broad
stoichiometric ordering of solid solutions, and nano-scale properties
within a bulk material or thin films (e.g. grain boundaries, domain
walls), simulation cells must contain thousands of atoms.
Both the shell model and the effective Hamiltonian have free 
parameters that are fitted from experimental and first-principles observations.
Fitting an isotropic shell model to first principles density functional theory
(DFT) calculations is generally easier than an effective model
Hamiltonian, which is much more complex. In addition a shell model
retains information on atomic positions.

The accuracy of a shell model depends on both the complexity of the
model and on the accuracy of the data used to fit the model, which for
DFT depends on approximations made in the exchange-correlation 
functional. Isotropic shell models fitted to results from DFT calculations have 
successfully described the qualitative behavior of ferroelectric
perovskites.
\cite{sepliarsky_atomic-level_2000,sepliarsky_atomistic_2004,tinte_ferroelectric_2004,
sepliarsky_atomic-level_2005,asthagiri_advances_2006,machado_phase_2010}
A shell model of BaTiO$_3$, fitted to DFT using the local density
approximation (LDA), significantly underestimates the lattice
constants, polarizations, and phase transition temperatures of
BaTiO$_3$.\cite{sepliarsky_atomistic_2004}.  Scaling of the lattice
constant to the experimental values does not correct issues with
the magnitude of polarization and phase transition
temperatures,\cite{tinte_atomistic_1999} and introduces
inconsistencies across different supercell sizes. 
The generalized gradient approximation (GGA) within DFT as 
implemented by Perdew, Burke, and Ernzerhoff (PBE) \cite{perdew_generalized_1996}
improves the accuracy of the lattice constants of all phases and
energy differences between phases, but it severely overestimates the
tetragonal strain and volume of each unit cell.\cite{asthagiri_advances_2006,wu_comparing_2004}
The failure of the mentioned approximations is due to the sensitivity of perovskite
ferroelectric properties to pressure. Increasing the pressure significantly 
decreases the well depths in the potential energy surface 
\cite{cohen_origin_1992}.

To improve on the issues with LDA and GGA/PBE in DFT calculations of
perovskites, two other forms for the exchange-correlation functional have been considered: the 
weighted density approximation (WDA)
\cite{gunnarsson_descriptions_1979, gunnarsson_exchange_1976,  
gunnarsson_exchange_1977, alonso_nonlocal_1978} and the
Perdew-Burke-Ernzerhof generalized gradient  approximation for solids
(PBEsol).\cite{perdew_restoring_2008}  
WDA gives much better cubic lattice constants for BaTiO$_3$, but
it does not improve the volume and strain of the other phases compared
to LDA and PBE.\cite{wu_comparing_2004}
The modification of PBE for solids (PBEsol) leads to very accurate lattice parameters for
the ground state,\cite{wu_more_2006} it slightly underestimates the cubic lattice 
constant and cuts the percentage error to one-third relative to
LDA and PBE. PBEsol does overestimate the $c/a$ ratio of the
tetragonal phase but less than PBE.\cite{wahl_srtio_3_2008} PBEsol
also gives much better oxygen-titanium displacements in {BaTiO$_3$}. 

In this paper we will discuss an isotropic shell model for {BaTiO$_3$} fitted
to results from DFT calculations using PBEsol.
{BaTiO$_3$} is chosen, since it has a rich phase diagram
and it has been used frequently for shell models, which allows for comparisons.
\cite{tinte_atomistic_1999, tinte_ferroelectric_2004,
  sepliarsky_atomic-level_2005, chen_modification_2009,
  sepliarsky_dynamical_2009, hu_a_2010} 
In section \ref{theory} the shell model is discussed as well as the methods behind the 
fitting procedure and DFT calculations.  In section \ref{results} the
result of our model are presented and compared to both the results of
{it ab initio} DFT calculations and experimental data for bulk
properties at finite temperatures.
Our conclusions are given in section \ref{conclusion}.
\section{Theory and Methods}
\label{theory}
The isotropic shell model treats an atom as two coupled 
charged particles connected by a spring: a core, which holds the 
atomic mass of the atom, and a massless shell.  All cores
and shells interact by Coulomb's law 
except the core and shell of the same atom.  A short-range interaction
that interacts only between shells of different atoms describes both
electron cloud repulsion and van der Waals attraction. The short-range interaction chosen is
the Buckingham potential and has the form 
$V(r) = A\exp(-r/\rho) - C/r^6$. 
The coupled core and shell interact via an anharmonic spring
$V(r) = k_2 r^2/2 + k_4 r^4 / 24$.  
Hence, the free parameters in the shell model consist of the core and shell charges 
and the constants in the Buckingham and anharmonic spring potentials.
Overall charge neutrality adds one constraint for the charges.

The free parameters for the shell model are found through least
squares minimization of the energy differences in the potential energy
surface between the shell model (SM) and the DFT calculations (Eqn.~\ref{chisq}).  The perfect cubic structure was chosen as the reference structure.  
Minimization of Eqn.~\ref{chisq} was achieved using the 
Nelder-Mead Downhill Simplex Method \cite{press_numerical_2007}.
\begin{equation}\label{chisq}
\chi^2 = \sum_{i=1}^N \sigma_i^2 \left[ (E_{\mathrm{C,DFT}} - E_{i,\mathrm{DFT}} ) - ( E_{\mathrm{C,SM}} - E_{i,\mathrm{SM}} ) \right]^2
\end{equation}
2200 configurations of the cubic, tetragonal, orthorhombic, 
and rhombohedral phases of {BaTiO$_3$} were used for the fitting.
Configurations include randomly 
displaced atoms from their optimal positions as well as changes in volume and strain.  
To determine the shell model energy $E_\mathrm{SM}$, for each configuration using a
trial set of the shell model parameters, the core of each atom is
placed at the ionic position of the atom in the DFT configuration. The 
shell model energy $E_\mathrm{SM}$ is determined by relaxing the
shells. Configurations with energies lower than the cubic
phase and with energies close to any of the four phases of {BaTiO$_3$} are
emphasized using a weight factor, $\sigma_i$, to reproduce the
DFT lattice constants accuarately.

DFT calculations using the PBEsol exchange correlation functional were performed using 
the projector augmented plane wave (PAW) method as implemented in the 
Vienna ab-initio simulation package 
(VASP) \cite{bloechl_projector_1994,kresse_ultrasoft_1999}.  
The plane-wave cutoff was set to 600 eV and
Brillouin-zone integration was performed using a 6 $\times$ 6 $\times$ 6 
Monkhorst-Pack (MP) k-point mesh,\cite{monkhorst_special_1976} 
using the tetrahedron method.\cite{bloechl_improved_1994}
For each configuration, the constrained groundstate was found by
relaxing the geometry until the maximum force on all atoms was less than 0.01 eV/\AA.
The orthorhombic and tetragonal states were found by relaxing the geometry of the unit cell while
restricting the movement of the titanium atom relative to the barium atom to the [110] and [001] 
directions, respectively.
Convergence was tested by comparing the  calculated zero temperature
lattice constants and energies of all four phases of {BaTiO$_3$} with
calculations using a plane-wave cutoff of 800 eV and a 8 $\times$ 8
$\times$ 8 MP k-point mesh.  
Lattice constants and total energies were converged to better than
0.002 \AA~and 1 meV/atom respectively. Our calculated unit cell
parameters agree very well with previously reported
values.\cite{wu_more_2006, wahl_srtio_3_2008} 
\section{Results and Discussion}
\label{results}
\begin{figure}
\includegraphics[scale=0.65]{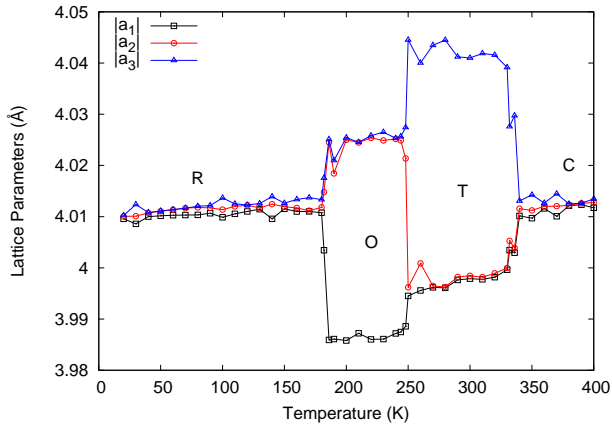}
\caption{Average lattice constants for all the three lattice vectors
  for {BaTiO$_3$} as a function of temperature determined from MD
  simulations using the shell model fitted to scaled energy
  differences (see text).}
\label{Scal_Lattice}
\end{figure}

\begin{figure}
\includegraphics[scale=0.65]{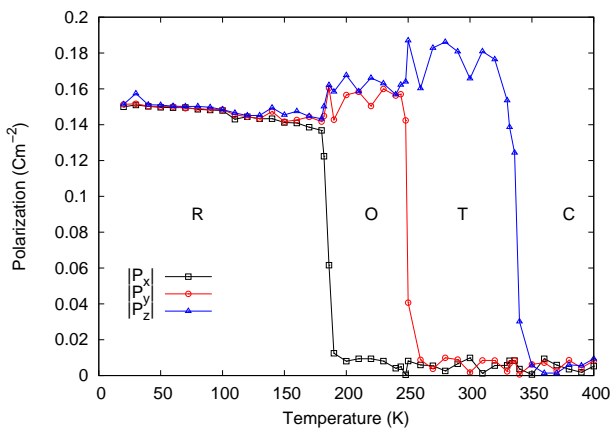}
\caption{Average absolute polarization in all three cartesian
  coordinates 
for {BaTiO$_3$} as a function of temperature determined from MD
  simulations using the shell model fitted to scaled energy
  differences (see text).}
\label{Scal_Polar}
\end{figure}

The shell model parameters derived for {BaTiO$_3$} from the fit of PBEsol
total energies are listed in Table \ref{parameters}.
Table \ref{parameters} lists 21 parameters, of which 16 were
determined in the fitting procedure. Effectively one core charge is
determined from charge neutrality.   
The repulsive $k_4$ parameters in the Buckhingham potential
for the cations were adapted from the work by Sepliarsky et
al\cite{sepliarsky_atomic-level_2005}. They help in the fitting
procedure, in particular by constraining the Ti shell, but they have
a negligible effect on the shell model energies.
The van der Waals interaction from the cations to the oxygen
atoms was set to zero because the separation between the core and the shell
of the cations is so small that they also have negligible effect on
the shell model energies. 
\begin{table}
\caption{Shell model parameters\footnote{Core and shell charges are in
    units of electrons energies in units of eV, and lengths in units
    of \AA.} 
    based on DFT total energies using the PBEsol exchange correlation 
    functional and determined by least square minimization (see
    text).} 
\label{parameters}
\begin{tabular}{|l|l|l|l|l|}  \hline\hline
Atom & Core charge & Shell charge & $k_2$ & $k_4$ \\ \hline
Ba & 4.859 & -2.948 & 311.64 & 0.0 \\ 
Ti   & 4.555 & -1.615 & 332.55 & 500.0 \\ 
O    & 1.058 & -2.675 & 46.31  & 6599.24 \\ \hline
Short-range & $A$ & $\rho$ & $C$ & \\ \hline
Ba-O & 1588.36 & 0.3553 & 0.0 & \\
Ti-O & 3131.25 & 0.2591 & 0.0 & \\ 
O-O  & 2641.41 & 0.3507 & 535.37 & \\  \hline\hline
\end{tabular}
\end{table}
Table~\ref{0Ktable} summarizes the zero temperature
properties for all four phases of {BaTiO$_3$} using the parameters from Table~\ref{parameters}.  
All lattice parameters for the shell model are within 0.01 \AA~of the
DFT lattice parameters. The  energy differences between the phases are
overestimated by the shell model by $\sim 2$ eV or 10\% relative to the DFT energy
differences (Tab. \ref{0Ktable}). 
\begin{table}
\caption{Shell model (SM) and scaled SM 0 K properties of all four
  phase of {BaTiO$_3$} and comparison to DFT results. $\Delta E$ is the
  energy of a phase relative to the cubic phase (see text).} 
\label{0Ktable}
\begin{tabular}{|l|l|l|l|l|} \hline \hline
  Phase &                       & DFT (PBEsol) & SM & Scaled SM\\ \hline
cubic   &   $a_o$ (\AA)         &    3.985 & 3.985 & 3.983\\ \hline
        &   $\Delta E$ (meV)    &   18.8 & 20.7 & 44.9\\ 
tetra   &   $a$ (\AA)           &   3.971 & 3.975 & 3.968\\
        &   $c/a$               &   1.023 & 1.022 & 1.028 \\ \hline
        &   $\Delta E$ (meV)    &   24.0 & 25.7 & 56.8\\ 
orth    &   $a$ (\AA)           &   3.964 & 3.971 & 3.961\\ 
        &   $c/a,b/a$           &   1.015 & 1.014 & 1.018\\ \hline
rhom    &   $\Delta E$ (meV)    &   25.3 & 27.8 & 59.4\\ 
        &   $a$ (\AA)           &   4.005 & 4.008 & 4.010\\ \hline \hline
\end{tabular}
\end{table}
Using the shell model the finite temperature behavior of {BaTiO$_3$} can be
determined form molecular dynamic (MD) simulations. 
All MD simulations were carried out using DL-POLY 
in a 10 $\times$ 10 $\times$ 10 supercell with 
periodic boundary conditions and variable cell shape in a (N, $\sigma$, T) 
ensemble at atmospheric pressure.\cite{smith_dl_1996}  
The barostat and thermostat relaxation times are set to 0.1 ps each. 
All shells are assigned a mass of 2 a.u, so that the
shell motion can be treated dynamically.\cite{asthagiri_advances_2006}
The time step used is 0.4 fs, the system is equilibrated for 4 ps,
followed by a simulation run for 12 ps.
\begin{figure}
\includegraphics[scale=0.65]{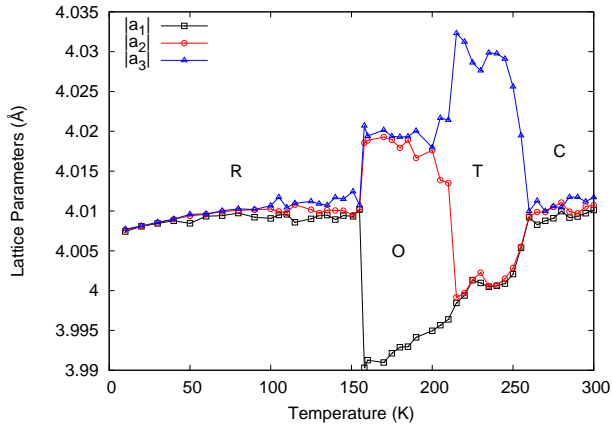}
\caption{Average lattice constants for all three lattice vectors
  for the {BaTiO$_3$} phases (cubic (C), tetragonal (T), orthorhombic (O),
  rhombohedral (R)) as a function of temperature determined from MD
  simulations using the shell model (see text).}
\label{Lattice}
\end{figure}
\begin{figure}
\includegraphics[scale=0.65]{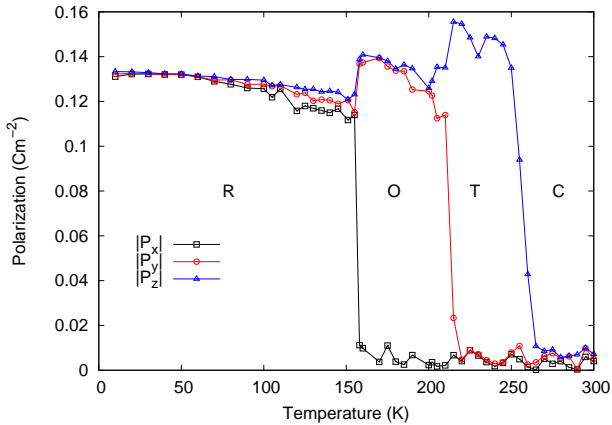}
\caption{Average absolute polarization in all three cartesian
  coordinates 
  for the {BaTiO$_3$} phases (cubic (C), tetragonal (T), orthorhombic (O),
  rhombohedral (R)) as a function of temperature determined from MD
  simulations using the shell model (see text).}
\label{Polar}
\end{figure}
The temperature dependent results of the MD simulations are shown in Figures \ref{Lattice}
and \ref{Polar}. The shell model does reproduce the correct 
sequence of phase transitions: Rhombohedral (0-150
K) $\rightarrow$ Orthorhombic (150 - 210 K)
$\rightarrow$ Tetragonal (210-260 K) $\rightarrow$ Cubic. The
experimental phase transitions occur at 
183, 278, and 393 K.  The PBEsol based shell model consistently underestimates the
temperatures of the phase transition temperatures, similar to previous
work using shell models fitted from DFT using
LDA,\cite{sepliarsky_atomistic_2004} and from DFT using LDA with
scaled lattice constants.\cite{tinte_atomistic_1999}
Averaged finite temperature lattice constants extracted from the MD
simulations are summarized in Table~\ref{phasetable} and with a
maximum difference of 0.01 \AA, agree very 
well with experimental values. The tetragonal and orthorhombic strain
are slightly underestimated. The calculated absolute
polarizations for the rhombohedral, orthorhombic, and tetragonal
phases are 0.225, 0.198, and 0.15 Cm$^{-2}$ respectively. These values
underestimate the experimental values, 0.33, 0.36, and 0.27
Cm$^{-2}$, by 55-68\%. This large disagreement is a result of
emphasizing energy differences between configurations in the
determination of the model parameters, while the polarization was not
part of the fitness function. In the shell model, both the
polarization and the total energy are very sensitive to the spring
interaction between the oxygen core and the shell.\cite{chen_modification_2009}
\begin{table}
\label{phasetable}
\caption{Finite temperature shell model lattice constants extracted
  from MD simulations for all four
  phase of {BaTiO$_3$} compared to experimental values.}
\begin{tabular}{|l|l|l|l|l|} \hline \hline
Phase        & & Shell Model & Scaled SM & Exp.\\ \hline
Cubic        & $a$ (\AA) & 4.01  & 4.01  & 4.00  \\ \hline
Tetragonal   & $a$ (\AA) & 4.00  & 4.00  & 3.995 \\ 
             & $c/a$     & 1.008 & 1.011 & 1.010 \\ \hline
Orthorhombic & $a$ (\AA) & 4.00  & 3.99  & 3.985 \\ 
             & $c/a,b/a$ & 1.006 & 1.010 & 1.008 \\ \hline 
Rhombohedral & $a$ (\AA) & 4.01  & 4.01  & 4.004 \\ \hline \hline
\end{tabular}
\end{table}
The phase transition temperatures can be improved by scaling the
differences in the potential energy surface in the DFT calculations,
which was successfully demonstrated using a shell model based on LDA
total energies and extrapolated lattice
constants.\cite{tinte_ferroelectric_2004}   
\begin{table}
\caption{Shell model parameters\footnote{Core 
 and shell charges are in units of electrons, 
 energies in units of eV, and lengths in units of \AA.}
 based on the scaled DFT total energies using the PBEsol exchange correlation
 functional and determined by least square minimization (see text).}
\label{parameters_scaled}
\begin{tabular}{|l|l|l|l|l|} \hline\hline
Atom & Core charge & Shell charge & $k_2$ & $k_4$ \\ \hline
Ba   & 5.042 & -2.870 & 298.51 & 0.0 \\ 
Ti   & 4.616 & -1.544 & 306.14 & 500.0 \\ 
O    & 0.970 & -2.718 & 36.93  & 5000.0 \\ \hline
Short-range & $A$ & $\rho$ & $C$ & \\ \hline
Ba-O & 7149.81 & 0.3019 & 0.0 & \\
Ti-O & 7200.27 & 0.2303 & 0.0 & \\ 
O-O  & 3719.60 & 0.3408 & 597.17 & \\ \hline\hline
\end{tabular}
\end{table}
 Rescaling the potential energy differences in order to reproduce the
rhombohedral-orthorhombic phase transition temperature results in the
shell model parameters listed in Table~\ref{parameters_scaled}. 
The MD simulation results for the scaled shell model are presented in
Figures \ref{Scal_Lattice} and \ref{Scal_Polar}. The calculated phase
transition temperatures using the scaled shell model are now 180 K, 250 K,
and 340 K, respectively, in much better agreement with experimental
values.  At the same time the lattice parameters of all 4 phases are
practically unchanged when compared to the values obtained with the
original, un-scaled shell model (Tables \ref{phasetable}, \ref{0Ktable}).
The calculated polarizations using the scaled shell model have
increased slightly relative to the first shell model for all 3
non-cubic phases, but the values are still much to small when compared
to experimental values.  
\section{Summary and Conclusions}
\label{conclusion}
We have developed and implemented an isotropic shell model from
fitting total energy differences in the potential energy surface
calculated using DFT with the PBEsol exchange correlation functional.  
Our model produces very accurate lattice constants when compared to
the experimental lattice constants at finite temperatures. The
sequence of phases of {BaTiO$_3$} is reproduced correctly but the transition
temperatures are too small.  A scaling of the 
potential energy surface results in much better agreement for the
phase transition temperatures, without sacrificing the accuracy of the
lattice constants. Using
total energies based on the PBEsol exchange correlation functional
offers significant benefits over LDA or GGA based energies even for
the basic isotropic shell model employed in this work. Various
enhancements provide room for further improvement, such as an
improved scaling of the energy differences, 
including ab-initio values of the polarization in the cost function,
introducing a distance dependent spring constant $k_2(r)$ for the
oxygen atoms\cite{chen_modification_2009}, and using an anisotropic 
spring constant for the oxygen core-shell system.
Ultimately, the PBEsol functional holds significant promise for the successful development
of shell models for more complex perovskite solid solutions entirely
from ab-inito total energy calculations.


%

\end{document}